# FELFCNCA: Fast & Efficient Log File Compression Using Non Linear Cellular Automata Classifier

P.Kiran Sree[*1], Inampudi Ramesh Babu[2], SSSN Usha Devi N[3]

[1]Research Scholar, Dept of CSE, JNTU Hyderabad, India ; [2] Professor, Dept of CSE, Acharya Nagarjuna University, India ; [3] Dept of CSE, BVCEC India.

[*1] profkiran@yahoo.com; [2] drirameshbabu@gmail.com; [3] usha.jntuk@gmail.com

*Abstract*

Log Files are created for Traffic Analysis, Maintenance, Software debugging, customer management at multiple places like System Services, User Monitoring Applications, Network servers, database management systems which must be kept for long periods of time. These Log files may grow to huge sizes in this complex systems and environments. For storage and convenience log files must be compressed. Most of the existing algorithms do not take temporal redundancy specific Log Files into consideration. We propose a Non Linear based Classifier which introduces a multidimensional log file compression scheme described in eight variants, differing in complexity and attained compression ratios. The FELFCNCA scheme introduces a transformation for log file whose compressible output is far better than general purpose algorithms. This proposed method was found lossless and fully automatic. It does not impose any constraint on the size of log file.

## Introduction

Log file keeps track of everything goes in and out of a particular server. It is a concept much like the black box of an airplane that records everything going on with the plane in the event of a problem. The information is frequently recorded chronologically, and is located in the root directory, or occasionally in a secondary folder, depending on how it is set up with the server. With the use of log file it's possible to get a good idea of where visitors are coming from, how often they return, and how they navigate through a site. Using cookies enables Webmasters to log even more detailed information about how individual users are accessing a site. The storage requirement for operational data is increasing as increase in the computing technology.

Current computer networks consist of a myriad of interconnected, heterogeneous devices in different physical locations (this is termed a distributed computing environment). Computer networks are used by many different sized corporations, educational institutions, small businesses and research organizations. In many of these networks, monitoring is required to determine the system state, improve security or gather information.

Log files are excellent sources for determining the health status of a system and are used to capture the events happened within an organization's system and networks. Logs are a collection of log entries and each entry contains information related to a specific event that has taken place within a system or network. Many logs within an association contain records associated with computer security which are generated by many sources, including operating systems on servers, workstations, networking equipment and other security software's, such as antivirus software, firewalls, intrusion detection and prevention systems and many other applications. Routine log analysis is beneficial for identifying security incidents, policy violations, fraudulent activity, and operational problems.

Initially, logs were used for troubleshooting problems, but nowadays they are used for many functions within most organizations and associations, such as optimizing system and network performance, recording the actions of users, and providing data useful for investigating malicious activity. Logs have evolved to contain information related to many different types of events occurring within networks and systems. Within an organization, many logs contain records related to computer security; common examples of these computer security logs are audit logs that track user authentication attempts and security device logs that record possible attacks.

## Log File Compression

This section motivates the need for data reduction and explains how data compression can be used to reduce the quantity of data. It also discusses the need for the approaches which can be used and the role of log files. In many environments, tracked logs can happen very often. As a result, there is a huge amount of data produced this way every day. And often it is necessary to store them for a long period of time. Regardless of the type of recorded logs, for reasons of simplicity and convenience, they are usually stored in plain text log files. Both the content type and the storage format suggest that it is possible to significantly reduce the size of log files through lossless data compression, especially if specialized algorithm was used. The smaller, compressed files have the advantages of being easier to handle and saving storage space.

## Types Of Compression

Lossless compression algorithms frequently use statistical redundancy in such a way as to represent the sender's data more concisely without error. Lossless compression is possible as most real-world data has statistical redundancy. For instance, in English text, the letter 'e' is much more used than the letter 'z', and the probability that the letter 'q' will be followed by the letter 'z' is very tiny. Another kind of compression called lossy data compression or perceptual coding is possible in this scheme some loss of data is acceptable. In general a lossy data compression will be guided by research on how people perceive the data in question. Forexample, the human eye is more sensitive to slight variations in luminance than it is to variations in color. JPEG image compression works in part by "rounding off" some of this less-important information.

## Lossy Compression

Lossy image compression is exercised in digital cameras, to increase storage capacity with negligible degradation of picture quality. Similarly, DVDs use the lossy MPEG-2 Video codec for compression. Compression of individual speech is often performed with more specialized methods, so that "speech compression" or "voice coding" is sometimes distinguished as a separate discipline from "audio compression". Different audio and speech compression standards are listed under audio codes. Voice compression is used in Internet telephony, for example while audio compression is used for CD ripping and is decoded by audio players.

Based on the literature reviews, it can be noted that many approaches can be considered to design and develop an intelligent system. For our system, we intend to implement RBR and OO methodology. RBR can be used to easily construct, debug and maintain the system and by using an Oracle database for knowledge representation, rules are more structured whereas OO methodology helps to structure the whole system in a conceptual view which will ease the programming implementation. The main objective of this study is to build a system for diabetes diagnosis based on RBR-OO methodology. To be precise, we aim to (i) design a general framework of the intelligent diagnostic system based on RBR-OO methodology, (ii) implement RBR algorithm to diagnose the patients based on the sign-symptoms provided, (iii) provide reasoning on the diagnosis and (iv) demonstrate the applicability of RBR-OO methodology in the healthcare environment.

## Non Linear Cellular Automata (CA)

Non Linear Cellular Automata use generalized structures to solve problems in an evolutionary way. CA often demonstrates also significant ability toward self-organization that comes mostly from the generalized structure on which they operate. By organization, one means that after some time in the evolutionary process, the system exhibits more or less stable generalized structures. This behaviour can be found no matter the initial conditions of the automaton. A CA consists of a number of cells organized in the form of a lattice.

## Genetic Algorithm & CA

The main motivation behind the evolving Non Linear Cellular Automata framework is to understand how genetic algorithms evolve Non Linear Cellular Automata that perform computational tasks requiring global information processing, since the individual cells in a CA can communicate only locally without the existence of a central control  the GA has to evolve CA that exhibit higher-level emergent behaviour in order to perform this global information processing. Thus this framework provides an approach to studying how evolution can create dynamical systems in which the interactions of simple components with local information storage and communication give rise to coordinated global information processing.

**Crossover and Mutation**

The performance of an individual is measured by a fitness function. A fitness function rates the performance of an individual in its environment by comparing the results of the individual's chromosomes with the desired results of the problem as defined by the author of the algorithm. The fitness is generally expressed within the algorithm as a floating point (i.e., decimal) number with a predefined range of values, from best performing to worst performing. As in Darwinian evolution, low-performing individuals are eliminated from the population and high performing individuals are cloned and mutated, replacing those that were eliminated.

The main feature which is achieved when developing CA Agent systems, if they work, is flexibility, since a CA Agent system can be added to, modified and reconstructed, without the need for detailed rewriting of the application.

**Emergent Computation In Non Linear Linear Cellular Automata**

A particular NLCA update rule and a corresponding initial configuration were constructed, or at least it was shown that such a construction is possible in principle. However, for most computational tasks of interest, it is very difficult to construct a NLCA update rule or it is very time consuming to set up the correct initial configuration.

A *genetic algorithm* (GA) is a search algorithm based on the principles of natural evolution (see the introduction to evolutionary computation in this volume). It NLCAn be used to search through the space of possible NLCA rules to try to find a NLCA that NLCAn perform a given computational task. The specific task that Mitchell *et al*. originally considered is *density classifiNLCAtion*. Consider one-dimensional, two-state NLCA. For any initial configuration (IC) of black and white cells, we NLCAn ask: "Are there more black cells or more white cells in the IC?" Note that for most "computing systems" (including humans), this is an easy question to answer, as the number of black and white cells NLCAn easily be counted and compared. However, for a NLCA this is a difficult task, since each individual cell in the lattice NLCAn only check the states of its direct neighbors, and none of the cells NLCAn have information of the global state of the NLCA lattice. So, the question is whether there exists a NLCA update rule that NLCAn decide on the density of black cells (smaller or larger than 0.5) for any IC.

Log files are plain text files in which every line corresponds to a single logged event description. The lines are separated by end-of-line marks. Each event description consists of at least several tokens, separated by spaces. A token may be a date, hour, IP address, or any string of characters being a part of event description. In typical log files the neighboring lines are very similar, not only in their structure, but also in their content. The proposed transform takes advantage of this fact by replacing the tokens of a new line with references to the previous line. There may be a row of more than one token that appears in two successive lines, and the tokens in the neighbouring lines can have common prefixes but different suffixes (the opposite is possible, but far less frequent in typical log files).

**Non Linear Cellular Automata Classifier**

NLCA Tree Building ( Assuming  K NLCA Basins)

Input:   Intrusion parameters

Output: NLCA Based inverted tree (LOG  INDEX)

Step 0: Start.

Step 1: Generate a NLCA with *k* number of NLCA basins

Step 2: Distribute the parameters into k NLCA basins

Step 3: Evaluate the distribution in each closet  basin

Step 4: Calculate the Rq

Step 5: Swap the more appropriate features to the bottom leaves of the inverted tree.

Step 6: Stop.

**Results & Discussion**

The result shows the two major indispensible properties of Log Files. The first one is the compressed file size and the other one is compaction ratio. Here the graphs show both of the properties. On one hand first graph shows the difference between original log files and compressed log files .On other hand second graph displays the compaction ratio. The third graph compares the existing method and the proposed method which shows that the proposed method is better than the existing method.

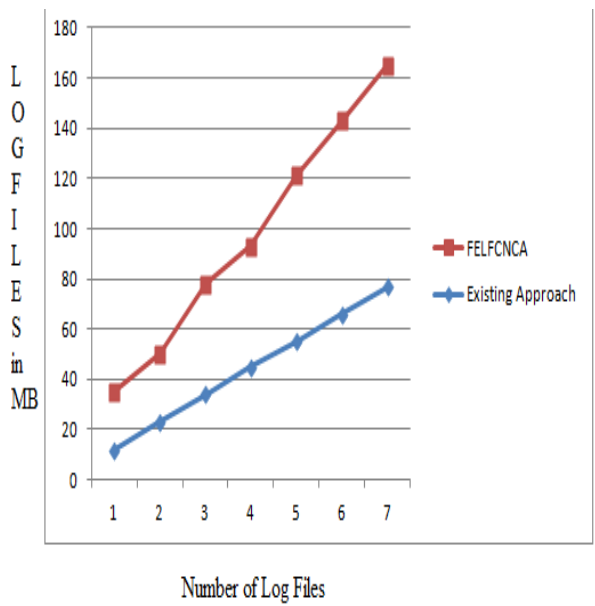

FIG. 1 COMPARISION AMONG FELFCNCA & EXISTING APPROACH

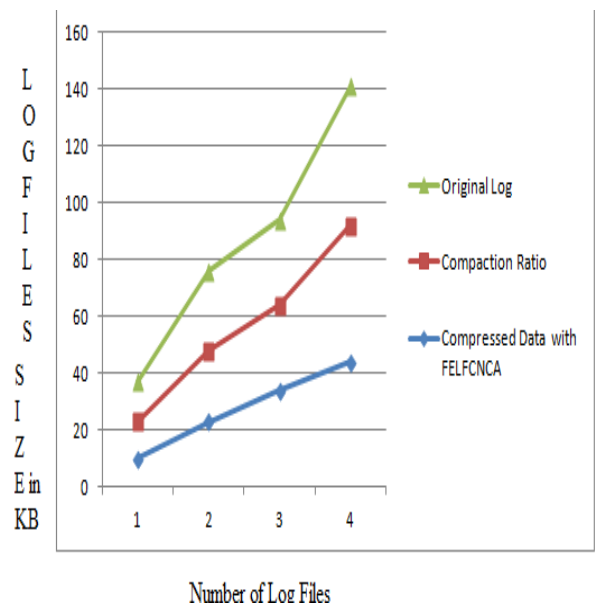

FIG. 2 COMPACTION RATIOS

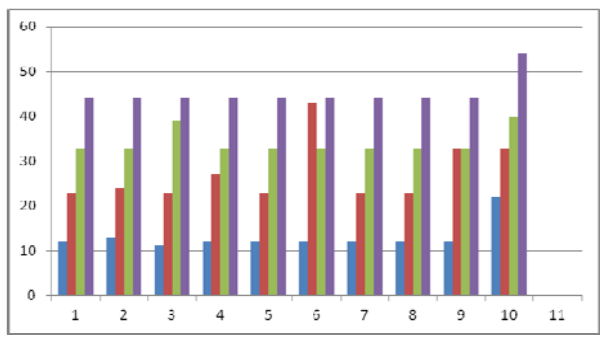

1.GZIP  2. GZIP+LP1  3. GZIP+LP2  4. GZIP+LP3  5. GZIP+LP3 6. LRZIP 7. BZIP2 8. LZMA 9. PPMVC 10.FCLFCNA

FIG 3 .COMPARISON OF COMPRESSION RATIOS

## Conclusion

System files contain lots of contemporary information replete with log files and occupy huge amounts of space. The general purpose algorithms do not take full advantage of redundancy in log files. The existing specialized compression techniques for log files needs lot of human assistance and not practical for the general applications. The proposed classifier introduces a multidimensional log file compression scheme described in eight variant differing in complexity and attained compression ratios. This FCLFCDNA classifier is designed for different types of logs and experimental results conclude the compression improvement in different log files. The proposed method is fully automatic and lossless. The most important functionality of this proposed work is it does no impose any constraints on the size of the log file.

P.Kiran Sree, I.Ramesh Bababu, "Identification of Protein Coding Regions in Genomic DNA Using Unsupervised FMACA Based Pattern Classifier" in International Journal of Computer Science & Network Security with ISSN: 1738-7906, Vol.8, No.1, 305-308.  TT

P.Kiran See, I.Ramesh Babu,Towards an Artificial Immune System to Identify and Strengthen Protein Coding Region Identification Using Cellular Automata Classifier, in International Journal of Computers and Communications, Issue 1, Volume 1, 2007,Issue 2, Volume 1, 2007,(26-34). ISSN: 2074-1294.

P.Kiran Sree, I. Ramesh Babu"Face Detection from still and Video Images using Unsupervised Cellular Automata with K means clustering algorithm" ICGST International Journal on Graphics, Vision and Image Processing(GVIP),ISSN: 1687-398X, Volume 8, Issue II, 2008, (1-7).

P.Kiran Sree, "Classification Scheme for IPv6 Routers Using Parallel Search of Individuals    Tries." in International Journal of Computer Science & Network Security with ISSN: 1738-7906, Vol. 8 No. 1  pp. 275-280

R. Lippmann, An introduction to computing with neural nets, IEEE ASSP Mag. 4(22) (2004), pp.121-129.

P. Maji and P. P. Chaudhuri (2004),"FMACA: A Fuzzy Cellular Automata Based  Pattern Classifier," Proceedings of 9th  International Conference on Database  Systems , Korea, 2004, pp. 494–505.

P. Maji and P. P. Chaudhuri, "Fuzzy Cellular Automata For Modeling Pattern Classifier," *Accepted for publication in IEICE,* (2004).

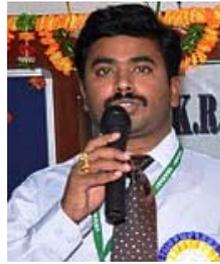

**P.KIRAN REE** received his B.Tech in Computer Science & Engineering, from J.N.T.U and M.E in Computer Science & Engineering from Anna University. He is pursuing Ph.D in Computer Science from J.N.T.U, Hyderabad. His areas of interests include Cellular Automata, Parallel Algorithms, Artificial Intelligence, and Compiler Design. He was the reviewer for some of International Journals and IEEE Society Conferences on Artificial Intelligence & Image Processing. He is also listed in Marquis Who's Who in the World, 29th Edition (2012), USA. Sree takes much commitment as the recipient of Bharat Excellence Award from Dr GV Krishna Murthy, Former Election Commissioner of India, and the Board of Studies member of Vikrama Simhapuri University, Nellore in Computer Science & Engineering stream; track chair for some international conferences. Meantime, he has published 30 technical papers both in international journals and conferences

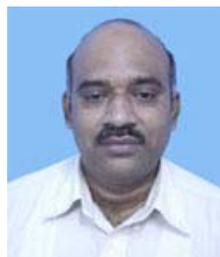

**Inampudi Ramesh Babu**, received his Ph.D in Computer Science from Acharya Nagarjuna University, M.E in Computer Engineering from Andhra University, B.E in Electronics & Communication Engg from University of Mysore. He is currently working as Head & Professor in the department of computer science, Nagarjuna University. Also he is the senate member of the same University from 2006. His areas of interest are image processing & its applications, and he is currently supervising 10 Ph.D students who are working in different areas of image processing. He is the senior member of IEEE, and  has published 70 papers in international conferences and journals.